\documentclass[onecolumn,showpacs,preprintnumbers,amsmath,amssymb]{revtex4}

\usepackage{color}
\usepackage{amsthm}
\usepackage{latexsym}
\usepackage{indentfirst}
\usepackage{graphicx}
\usepackage{placeins}
\usepackage{booktabs}
\usepackage{multirow}

\theoremstyle{plain}
\newtheorem{theorem}{Theorem}

\theoremstyle{definition}

\theoremstyle{remark}

\DeclareMathOperator{\tr}{Tr}

\def\angstrom{\ensuremath{\text{\AA}}}
\newcommand{\thmref}[1]{Theorem~\ref{#1}}

\newcommand{\secref}[1]{Section~\ref{#1}}
\newcommand{\figref}[1]{Figure~\ref{#1}}
\newcommand{\tabref}[1]{Table~\ref{#1}}

\newcommand{\etal}{\textit{et al}{}}
\newcommand{\ie}{\textit{i.e.}~{}}
\newcommand{\eg}{\textit{e.g.}~{}}

\newcommand{\ud}{\,\mathrm{d}}

\newcommand{\ZZ}{\mathbb{Z}}
\newcommand{\Or}{\mathcal{O}}

\newcommand{\bvec}[1]{\boldsymbol{\mathrm{#1}}}

\newcommand{\eps}{\epsilon}
\newcommand{\kB}{k_\mathrm{B}}
\newcommand{\eV}{\mathrm{eV}}
\newcommand{\K}{\mathrm{K}}

\newcommand{\abs}[1]{\lvert#1\rvert}

\renewcommand{\Re}{\mathrm{Re}~}
\renewcommand{\Im}{\mathrm{Im}~}

\begin{document}

\title{Multipole Representation of the Fermi Operator  
with Application to
the Electronic Structure Analysis of Metallic Systems}

\author{Lin Lin}
\affiliation{Program in Applied and Computational Mathematics, 
Princeton University, Princeton, NJ 08544}
\author{Jianfeng Lu}
\affiliation{Program in Applied and Computational Mathematics, 
Princeton University, Princeton, NJ 08544}
\author{Roberto Car}
\affiliation{Department of Chemistry and Princeton Center for Theoretical Science, Princeton University, Princeton, 
NJ 08544}
\author{Weinan E}
\affiliation{Department of Mathematics and PACM, Princeton University, 
Princeton, NJ 08544}

\begin{abstract}
We propose a multipole representation of the Fermi-Dirac function
and the Fermi operator,
and use  this representation to develop
algorithms for electronic structure analysis of
metallic systems. 
The new algorithm is quite simple and efficient. Its computational cost
scales logarithmically with $\beta\Delta\eps$ where $\beta$ is the inverse
temperature, and $\Delta \eps$ is the width of the spectrum of the 
discretized Hamiltonian matrix.
\end{abstract}

\maketitle

\section{Introduction}

The Fermi operator, \ie the Fermi-Dirac function of the system Hamiltonian,
is a fundamental quantity in the quantum mechanics of many-electron systems
and is ubiquitous in condensed matter physics. In the last decade the
development of accurate and numerically efficient representations of the Fermi
operator has attracted lot of attention in the quest for linear scaling
electronic structure methods based on effective one-electron Hamiltonians.
These approaches have numerical cost that scales linearly with $N$, the number
of electrons, and thus hold the promise of making quantum mechanical
calculations of large systems feasible. Achieving linear scaling in realistic
calculations is very challenging.  Formulations
based on the Fermi operator are appealing because this operator gives directly
the single particle density matrix without the need for Hamiltonian
diagonalization. At finite temperature the density matrix can be expanded in
terms of finite powers of the Hamiltonian, requiring computations that scale
linearly with $N$ owing to the sparse character of the Hamiltonian
matrix\footnote{When the effective Hamiltonian depends on the density, such as
\eg in density functional theory, the number of iterations needed to achieve
self-consistency may be an additional source of size dependence. This issue
has received little attention so far in the literature and will not be
considered in this paper.}. These properties of the Fermi operator are valid
not only for insulators but also for metals, making formulations based on the
Fermi operator particularly attractive.

Electronic structure algorithms using a Fermi operator expansion (FOE) were
introduced by Baroni and Giannozzi \cite{BaroniGiannozzi1992} and Goedecker
and co-workers \cite{goedeckercolombo1994, goedeckerteter1995} (see also the 
review article \cite{goedecker1999}). 
These authors
proposed polynomial and rational approximations of the Fermi operator. Major
improvements were made recently in a series of publications by Parrinello and
co-authors \cite{parrinello2005, parrinello2006a, parrinello2006b,
parrinello2007, parrinello2008, parrinelloarxiv}, in which a new form of Fermi
operator expansion was introduced based on the grand canonical formalism.     

From the viewpoint of efficiency, a major concern is the cost of the representations of the Fermi operator as a function of $\beta \Delta \eps$ where $\beta$ is the inverse temperature and $\Delta\eps$ is the spectral width of the Hamiltonian matrix. The cost of the original FOE proposed by 
Goedecker \etal\ scales as $\beta \Delta \eps$. The fast polynomial summation
technique introduced by Head-Gordon \etal \cite{headgordon2003,headgordon2004}
reduces the cost to $(\beta\Delta\eps)^{1/2}$. The cost of the hybrid
algorithm proposed by Parrinello \etal\ in a recent preprint
\cite{parrinelloarxiv} scales as $(\beta\Delta\eps)^{1/3}$.

The main purpose of this paper is to present a strategy that reduces the cost
to logarithmic scaling $(\ln\beta\Delta\eps)\cdot (\ln\ln \beta\Delta\eps)$,
thus greatly improving the efficiency and accuracy of numerical FOEs. Our
approach is based on the exact pole expansion of the Fermi-Dirac function which underlies the Matsubara formalism of finite temperature Green's functions in many-body physics \cite{mahan2000mpp}. It is natural to consider a multipole representation of this expansion to achieve better efficiency, as was done in the fast multipole method \cite{GreengardRokhlin1987}. Indeed, as we will show below, the multipole expansion that we propose does achieve logarithmic scaling. We believe that this representation will be quite useful both as a theoretical tool and as a starting point for computations. As an application of the new formalism, we present an algorithm for electronic structure calculation that has the potential to become an efficient linear scaling algorithm for metallic systems.

The remaining of the paper is organized as follows. In the next section, we
introduce the multipole representation of Fermi operator.
In \secref{sec:algorithm}, we present the FOE algorithm based on the multipole representation and analyze its computational cost. 
Three examples illustrating the algorithm are discussed in 
\secref{sec:example}. We conclude the paper with some remarks on future directions.

\section{Multipole representation for the Fermi operator}
\label{sec:FMM}

Given the effective one-particle Hamiltonian $\bvec{H}$, the inverse temperature $\beta = 1/\kB T$ and the chemical potential $\mu$, the finite temperature single-particle density matrix of the system is given by the Fermi operator  
\begin{equation}\label{eq:FDdist}
\bvec{\rho} = \frac{2}{1 + \exp(\beta(\bvec{H}-\mu))} 
= 1 - \tanh\Bigl(\frac{\beta}{2}(\bvec{H}-\mu)\Bigr),
\end{equation}
where $\tanh$ is the hyperbolic tangent function.
The Matsubara representation of the Fermi-Dirac function is given by 
\begin{equation}\label{eq:FOE}
\bvec{\rho} = 1 - 4\Re \sum_{l=1}^{\infty} \frac{1}{\beta(\bvec{H}-\mu) 
-(2l-1)\pi i}.
\end{equation}
This representation originates from the pole expansion 
(see, for example, \cite{JeffreysJeffreys, Ahlfors})
of the meromorphic function $\tanh$
\begin{equation}\label{eq:poleexpnc}
\tanh z = \sum_{l=-\infty}^{\infty} \frac{1}{z-\frac{2l-1}{2}\pi i}.
\end{equation}
In particular, for $x$ real (which is the case when $\bvec{H}$ is self-adjoint), we have
\begin{equation}\label{eq:poleexpnr}
\tanh x = 4\Re \sum_{l=1}^{\infty} \frac{1}{2x - (2l-1)\pi i}.
\end{equation}
To make the paper self-contained, we provide in Appendix A a simple derivation of this representation. It should be emphasized that \eqref{eq:FOE} is \emph{exact}. We notice that the expansion \eqref{eq:FOE} can also be understood as the $P\to\infty$ limit of an exact Fermi operator expansion proposed by Parrinello and co-authors in 
\cite{parrinello2005, parrinello2006a, parrinello2006b, parrinello2007, 
parrinello2008, parrinelloarxiv}.

\begin{figure}[ht]
	\begin{center}
		\includegraphics{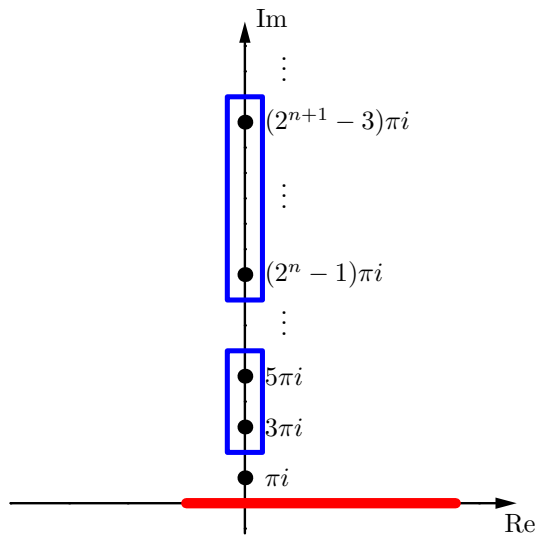}
	\end{center}
	\caption{(color online) Illustration of the pole decomposition
	\eqref{eq:polesumfmm}.  From $2^{n}$ to $2^{n+1}-1$ poles are grouped
	together as shown in the figure. The spectrum is indicated by the
	red line on the real axis.}
	\label{fig:poledecomp}
\end{figure}

The summation in \eqref{eq:FOE} can be seen as a summation of residues contributed from the poles 
$\{(2l-1)\pi i\}$, with $l$ a positive integer, on the imaginary axis. 
This suggests to look for a multipole expansion of the contributions from the poles, as done in the fast multipole method (FMM) \cite{GreengardRokhlin1987}.
To do so, we use a dyadic decomposition of the poles, in which the $n$-th
group contains terms from $l=2^{n-1}$ to $l=2^{n}-1$, for a total of $2^{n-1}$
terms (see \figref{fig:poledecomp} for illustration). We decompose the
summation in Eq.\eqref{eq:FOE} accordingly, denoting $x = \beta(\bvec{H}-\mu)$
for simplicity 
\begin{equation}
\sum_{l=1}^{\infty} \frac{1}{x-(2l-1)\pi i}
=\sum_{n=1}^{\infty}\sum_{l=2^{n-1}}^{2^{n}-1} \frac{1}{x-(2l-1)\pi i}
=\sum_{n=1}^{\infty} S_n.
\end{equation}
The basic idea is to combine the simple poles into a set of multipoles at $l =
l_n$, where $l_n$ is taken as the midpoint of the interval $[2^{n-1},2^n-1]$
\begin{equation}
l_n = \frac{3\cdot 2^{n-1}-1}{2}.
\end{equation}
Then the $S_n$ term in the above equation can be written as
\begin{equation}\label{eq:Sn}
\begin{aligned}
S_n &= \sum_{l=2^{n-1}}^{2^{n}-1}\frac{1}{x-(2l_n-1)\pi i - 2(l-l_n)\pi i}\\
&= \sum_{l=2^{n-1}}^{2^{n}-1}\frac{1}{x-(2l_n-1)\pi i} \sum_{\nu=0}^{\infty}
\Bigl(\frac{2(l-l_n)\pi i}{x-(2l_n-1)\pi i}\Bigr)^{\nu} \\
&= \sum_{l=2^{n-1}}^{2^{n}-1}\frac{1}{x-(2l_n-1)\pi i} \sum_{\nu=0}^{P-1}
\Bigl(\frac{2(l-l_n)\pi i}{x-(2l_n-1)\pi i}\Bigr)^{\nu} \\
&\qquad\qquad +\sum_{l=2^{n-1}}^{2^{n}-1}
\frac{1}{x-(2l-1)\pi i}\Bigl(\frac{2(l-l_n)\pi i}{x-(2l_n-1)\pi i}
\Bigr)^P
\end{aligned}
\end{equation}
In deriving Eq. \eqref{eq:Sn} we used the result for the sum of a geometric series. Using the fact that $x$ is real, the second term in 
Eq. \eqref{eq:Sn} can be bounded by
\begin{equation}
\sum_{l=2^{n-1}}^{2^{n}-1} \Bigl\lvert\frac{1}{x-(2l-1)\pi i}\Bigr\rvert
\Bigl\lvert \frac{2(l-l_n)\pi i}{x-(2l_n-1)\pi i}
\Bigr\rvert^P 
\leq \sum_{l=2^{n-1}}^{2^{n}-1} \frac{1}{\lvert (2l-1)\pi\rvert}
\Bigl\lvert\frac{2(l-l_n)}{2l_n-1}\Bigr\rvert^P \leq \frac{1}{2\pi} 
\frac{1}{3^P}
\end{equation}
Therefore, we can approximate the sum $S_n$ by the first $P$ terms, and the error decays exponentially with $P$:
\begin{equation}
\label{eq:errorP}
\left\lvert S_n(x) - 
\sum_{l=2^{n-1}}^{2^{n}-1}\frac{1}{x-(2l_n-1)\pi i} \sum_{\nu=0}^{P-1}
\Bigl(\frac{2(l-l_n)\pi i}{x-(2l_n-1)\pi i}\Bigr)^{\nu}\right\rvert \leq 
\frac{1}{2\pi}\frac{1}{3^P},
\end{equation}
uniformly in $x$. The above analysis is of course standard from the view point
of the fast multipole method \cite{GreengardRokhlin1987}. 
The overall philosophy is also similar: given a preset error tolerance, one selects $P$, the number of terms to retain in $S_n$, according to Eq. \eqref{eq:errorP}. 

\begin{figure}[ht]
	\begin{center}
		\includegraphics[width=7cm]{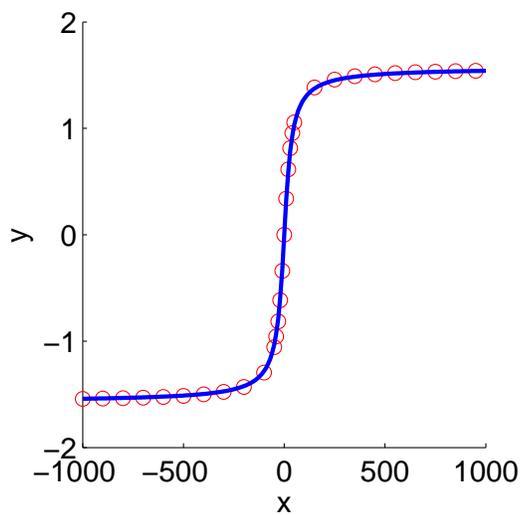}
	\end{center}
	\caption{(color online) The function $\Im
	\psi\left(m-\frac{1}{2}+\frac{i}{\pi}x\right)$(red circle), \ie the remainder of the pole expansion in
	Eq.~\eqref{eq:polesumfmm} is compared  
	with the function $\arctan\left(\frac{2x}{(2m-1)\pi}\right)$ (blue solid line) for $m=10$}
	\label{fig:psiatan}
\end{figure}

Interestingly, the remainder of the sum in Eq. \eqref{eq:FOE} from $l=m$ to $\infty$ has an explicit expression
\begin{equation}\label{eq:digamma}
\Re\sum_{l=m}^{\infty}\dfrac{1}{2x-(2l-1)i\pi} = \frac{1}{2\pi}\Im 
\psi\left(m-\frac{1}{2}+\frac{i}{\pi}x\right),
\end{equation}
where $\psi$ is the digamma function 
$\psi(z) = \Gamma'(z)/\Gamma(z)$. It is well known \cite{JeffreysJeffreys} 
that the digamma function has the following asymptotic expansion 
\begin{equation}
\psi(z) \sim \ln(z) - \frac{1}{2z} -
\frac{1}{12z^2}+\Or\Bigl(\frac{1}{z^4}\Bigr),
\quad \abs{\arg z}\leq\pi \text{ and } \abs{z}\to\infty.
\end{equation}
Therefore, 
\begin{equation}
\begin{aligned}
\Im \psi\left(m-\frac{1}{2}+\frac{i}{\pi}x\right) 
& \sim \Im\ln\left(m-\frac{1}{2}-\frac{i}{\pi}x\right) 
+ \Or\Bigl(\frac{1}{m^2}\Bigr) \\
& = \arctan\left(\frac{2x}{(2m-1)\pi}\right) + \Or\Bigl(\frac{1}{m^2}\Bigr)
, \quad m \to \infty.
\end{aligned}
\label{eq:psiasympt}
\end{equation}
\figref{fig:psiatan} shows that the asymptotic approximation
\eqref{eq:psiasympt} is already rather accurate when $m=10$.

Eq.~\eqref{eq:psiasympt}~also shows the effectiveness of the multipole representation from the viewpoint of traditional polynomial approximations. 
At zero temperature, the Fermi-Dirac function is a step function that cannot
be accurately approximated by any finite order polynomial. At finite but low
temperature, it is a continuous function with a very large derivative at
$x=0$, \ie when the energy equals the chemical potential $\mu$. The magnitude of this derivative becomes smaller and, correspondingly, the Fermi function becomes smoother as the temperature is raised. One can use the value of the derivative of the Fermi function at $x=0$ to measure the difficulty of an FOE. After eliminating the first $m$ terms in the expansion, Eq.~\eqref{eq:psiasympt} shows that asymptotically the derivative is multiplied by the factor $\frac{2}{(2m-1)\pi}$, which is equivalent to a rescaling of the temperature by the same factor. In particular, if we explicitly include the first $2^N$ terms in the multipole representation of the Fermi operator, we are left with a remainder which is well approximated by Eq.~\eqref{eq:psiasympt}, so that, effectively, the difficulty is reduced by a factor $2^{N}$. As a matter of fact standard polynomials approximations, such as the Chebyshev expansion, can be used to efficiently represent the remainder in Eq. \eqref{eq:digamma} even at very low temperature.

In summary, we arrive at the following multipole representation for the Fermi operator 
\begin{multline}\label{eq:polesumfmm}
\bvec{\rho} = 1 - 4 \Re\sum_{n=1}^{N}
\sum_{l=2^{n-1}}^{2^{n}-1}\frac{1}{\beta(\bvec{H}-\mu)-(2l_n-1)\pi i} 
\sum_{\nu=0}^{P-1}
\Bigl(\frac{2(l-l_n)\pi i}{\beta(\bvec{H}-\mu)-(2l_n-1)\pi i}\Bigr)^{\nu} \\ -\frac{2}{\pi}\Im\psi\left(2^N-\frac{1}{2}+\frac{i}{2\pi}\beta(\bvec{H}-\mu)\right)
+\Or(N/3^{P}).
\end{multline}
The multipole part is evaluated directly as discussed below, and the remainder is evaluated with the standard polynomial method.

\section{Numerical calculation and error analysis} 
\label{sec:algorithm}

To show the power of the multipole expansion, we discuss a possible algorithm to compute the Fermi operator in electronic structure calculations and present a detailed analysis of its cost in terms of $\beta \Delta \eps$.  
Given the Hamiltonian matrix $\bvec{H}$, it is straightforward to compute the density matrix $\bvec{\rho}$ from the multipole expansion if we can
calculate the Green's functions
$\bvec{B}_{l_n}=\left[\beta(\bvec{H}-\mu)-(2l_n-1)\pi i\right]^{-1}$ for different $n$. 

A possible way to calculate the inverse matrices is by the Newton-Schulz iteration. For any non-degenerate matrix $\bvec{A}$, the
Newton-Schulz iteration computes the inverse $\bvec{B}=\bvec{A}^{-1}$ as
\begin{equation}
	\bvec{B}_{k+1}=2\bvec{B}_{k}-\bvec{B}_{k}\bvec{A}\bvec{B}_k.
	\label{}
\end{equation}
The iteration error is measured by the spectral radius, \ie the eigenvalue of
largest magnitude, of the matrix $\bvec{I}-\bvec{B}\bvec{B}_{k}$ where
$\bvec{I}$ is the identity matrix. In the following we denote the spectral
radius of the matrix $\bvec{A}$ by $\sigma(\bvec{A})$. Then the spectral
radius at the $k$-th step of the Newton-Schulz iteration is
$\bvec{R}_k=\sigma(\bvec{I}-\bvec{B}\bvec{B}_{k})$ and 
\begin{equation}\label{eq:newtonerror} \sigma(\bvec{R}_{k+1}) =
	\sigma(\bvec{R}_{k})^2 = \sigma(\bvec{R}_0)^{2^{k+1}}.  \end{equation}
Thus the Newton-Schulz iteration has quadratic convergence.  
With a proper choice of the initial guess (see \cite{parrinello2008}), the number of iterations required to converge is bounded by a constant, and this constant depends only on the target accuracy.

The remainder, \ie the term associated to the
digamma function in Eq.~\eqref{eq:polesumfmm}, can be evaluated by 
standard polynomial approximations such as the Chebyshev expansion. The order of Chebyshev polynomials needed for a given target accuracy is proportional to
$\beta\Delta\eps/2^{N+1}$ (see \cite[Appendix]{headgordon1997}).  

Except for the error coming from the truncated multipole representation, the main source of error in applications comes from the numerical approximation of the Green's functions $\bvec{B}_{l_n}$. To understand the impact of this numerical error on the representation of the Fermi operator, let us rewrite
\begin{equation*}
\bvec{S}_n=\sum_{l=2^{n-1}}^{2^{n}-1} \bvec{B}_{l_n} 
\sum_{\nu=0}^{P-1} (-2(l-l_n)\pi i\bvec{B}_{l_n})^{\nu}
=\sum_{\nu=0}^{P-1} \bvec{B}_{l_n}^{\nu+1} 
\sum_{l=2^{n-1}}^{2^{n}-1} (-2(l-l_n)\pi i)^{\nu}.
\end{equation*}
The factor $\sum_l (-2(l-l_n)\pi i)^{\nu}$ is large, but we can control the total error in $S_n$ in terms of the spectral radius  
$\sigma(\bvec{B}_{l_n} -\widehat{\bvec{B}}_{l_n})$. Here
$\widehat{\bvec{B}}_{l_n}$ is the numerical estimate of $\bvec{B}_{l_n}$.

The error is bounded by 
\begin{equation}
\sigma(\widehat{\bvec{S}}_n - \bvec{S}_n) \leq \sum_{\nu=0}^{P-1}
2^{n-1}(2^{n-1}\pi)^{\nu}\sigma
\bigl(\bvec{B}^{\nu+1} - \widehat{\bvec{B}}^{\nu+1}\bigr)
\leq \sum_{\nu=0}^{P-1} (2^{n-1}\pi)^{\nu+1}\sigma
\bigl(\bvec{B}^{\nu+1} - \widehat{\bvec{B}}^{\nu+1}\bigr),
\end{equation}
where we have omitted the subscript $l_n$ in $\bvec{B}_{l_n}$ and in
$\widehat{\bvec{B}}_{l_n}$. In what follows the quantity $\sum_{\nu=0}^{P-1}
(2^{n-1}\pi)^{\nu+1}\sigma
\bigl(\bvec{B}^{\nu+1} - \widehat{\bvec{B}}^{\nu+1}\bigr)$ will be denoted by $e_P$.  
Then we have
\begin{equation}
\begin{aligned}
e_P & = \sum_{\nu=0}^{P-1} (2^{n-1}\pi)^{\nu+1}\sigma
\bigl((\bvec{B}^{\nu}-\widehat{\bvec{B}}^{\nu})\bvec{B} 
+ (\widehat{\bvec{B}}^{\nu} -\bvec{B}^{\nu})
(\bvec{B}-\widehat{\bvec{B}}) 
+ \bvec{B}^{\nu}(\bvec{B}-\widehat{\bvec{B}})\bigr) \\
& \leq \sum_{\nu=1}^{P-1} (2^{n-1}\pi)^{\nu+1}\bigl(\sigma(\bvec{B}) +
\sigma(\bvec{B}-\widehat{\bvec{B}})\bigl)
\sigma
(\bvec{B}^{\nu}-\widehat{\bvec{B}}^{\nu})+
\sum_{\nu=0}^{P-1}(2^{n-1}\pi)^{\nu+1}\sigma(\bvec{B})^{\nu}\sigma(\bvec{B}-\widehat{\bvec{B}}).\\
\end{aligned}
\end{equation}
Here we took into account the fact that the $\nu=0$ term in the first summation is equal to zero, and have used the properties $\sigma(\bvec{A}+\bvec{B})\leq \sigma(\bvec{A})+\sigma(\bvec{B})$, and
$\sigma(\bvec{A}\bvec{B})\leq\sigma(\bvec{A})\sigma(\bvec{B})$, respectively.

Noting that $2^{n-1}\pi\sigma(\bvec{B}_{l_n}) \leq 1/3$ and changing $\nu$ to
$\nu+1$ in the first summation,  we can rewrite $e_P$ as
\begin{equation}
\begin{aligned}
e_P & \leq \bigl(\frac{1}{3} + 2^{n-1}\pi\sigma(\bvec{B}-\widehat{\bvec{B}})\bigr)
\sum_{\nu=0}^{P-2}(2^{n-1}\pi)^{\nu+1}
\sigma(\bvec{B}^{\nu+1}-\widehat{\bvec{B}}^{\nu+1})
+ \sum_{\nu=0}^{P-1}\frac{1}{3^{\nu}} 2^{n-1}\pi
\sigma(\bvec{B}-\widehat{\bvec{B}}) \\
& \leq (\frac{1}{3} + 2^{n-1}\pi\sigma(\bvec{B}-\widehat{\bvec{B}}))e_{P-1} 
+ \frac{3}{2}2^{n-1}\pi\sigma(\bvec{B}-\widehat{\bvec{B}})\\
& = (\frac{1}{3} + e_1)e_{P-1} 
+ \frac{3}{2}e_1.\\
\end{aligned}
\end{equation}
In the last equality, we used the fact that
$e_1 = 2^{n-1}\pi\sigma(\bvec{B}-\widehat{\bvec{B}})$.
Therefore, the error $e_P$ satisfies the following recursion formula
\begin{equation}
e_P + \frac{3e_1/2}{e_1 - 2/3}\leq(\frac{1}{3} + e_1) \left(e_1 + 
\frac{3e_1/2}{e_{P-1} - 2/3}\right) 
\leq (\frac{1}{3} + e_1)^{P-1} \left(e_1 + 
\frac{3e_1/2}{e_1 - 2/3}\right).
\label{eq:recursion}
\end{equation}

Taking $e_1\leq \frac{2}{3}$, we have 
\begin{equation}
e_P\leq e_1 = 2^{n-1}\pi\sigma(\bvec{B}-\widehat{\bvec{B}}).
\end{equation}

Therefore, using Eq.~\eqref{eq:newtonerror} we find that the number $k$ of Newton-Schulz iterations must be bounded as dictated by the following inequality in order for the error $\sigma(\widehat{\bvec{S}}_n - \bvec{S}_n)$ to be $\leq 10^{-D}/N$.
\begin{equation}
	2^{n-1}\sigma(\bvec{R}_{0})^{2^k}\leq \frac{10^{-D}}{N}.
	\label{}
\end{equation}
Here we have used the fact that $\sigma(\bvec{B}_{l_{n}})\leq 1/\pi$ for any
$n$.  Each Newton-Schulz iteration requires two matrix by matrix multiplications, and the number of matrix by matrix multiplications needed in the Newton-Schulz iteration for $\bvec{B}_{l_n}$ with $n<N$ is bounded by 
\begin{equation}
2\log_2\left(\frac{D\log_2 10 + N + \log_2 N }{-\log_2
\sigma(\bvec{R}_0)}\right).
\end{equation}

To obtain a target accuracy $\sigma(\bvec{\rho}-\widehat{\bvec{\rho}})\leq
10^{-D}$ for a numerical estimate $\widehat{\bvec{\rho}}$ of the density matrix, taking into account the operational cost of calculating the remainder and the direct multipole summation in the FOE, the number of matrix by matrix multiplications $n_{\mathrm{MM}}$ is bounded by
\begin{equation}\label{eq:nmmmdetail}
n_{\mathrm{MM}} 
\leq 2 N\log_2 N + C_1 N + C_2 2^{-N-1}\beta\Delta\eps.
\end{equation}
Here we used the property: $\log_2 (x+y)\leq \log_2 x+\log_2 y$ when $x,y
\geq 2$, and defined the constant $C_1$ as follows:
\begin{equation}
	C_1=\frac{2}{N}\sum_{n=1}^{N}\log_{2}\left( 
	\frac{D\log_{2}10+\log_2 N}{-\log_2 \sigma\bigl( (\bvec{R}_0)_{l_n} \bigl)}
	\right).
	\label{}
\end{equation}
The dependence on $2^{-N-1}\beta\Delta\eps$ in the last term on the right hand
side of \eqref{eq:nmmmdetail} comes from Chebyshev expansion used to calculate the remainder. From numerical calculations on model systems, the constant $C_1$ and $C_2$ will be shown to be rather small.
Finally, choosing $N \propto \ln(\beta\Delta\eps)$,  we obtain
\begin{equation}\label{eq:nmmloge}
n_{\mathrm{MM}} \propto (\ln\beta\Delta\eps)\cdot(\ln\ln\beta\Delta\eps)
\end{equation}
with a small prefactor. 

\section{Numerical Examples}
\label{sec:example}

We illustrate the algorithm in three simple cases. 
The first is an off-lattice one dimensional model defined in a supercell with periodic boundary conditions. In this example, we discretize the Hamiltonian with the finite difference method, resulting in a very broad spectrum with a width of about $2000\eV$, and we choose a temperature as low as $32\K$. In the second example we consider a nearest neighbor tight binding Hamiltonian in a three dimensional simple cubic lattice and set the temperature to $100\K$. In the third example we consider a three dimensional Anderson model with random on-site energy on a simple cubic lattice at $100\K$. 

\subsection{One dimensional model with large spectral width}\label{sec:TestA}
 
\begin{figure}[ht]
	\begin{center}
		\includegraphics[scale=0.7]{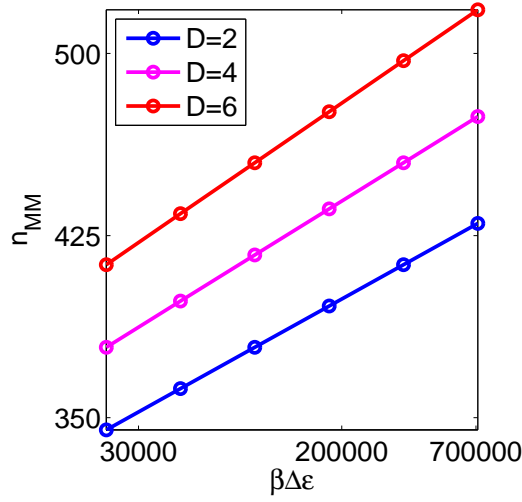}
	\end{center}
	\caption{(color online) Linear-log plot of the number of matrix matrix multiplications 
	$n_{\mathrm{MM}}$ versus $\beta\Delta\eps$.
	$n_{\mathrm{MM}}$ depends logarithmically on $\beta\Delta\eps$ with a small
	constant prefactor.}
	\label{fig:nmm1dfmm}
\end{figure}

In this example, a one dimensional crystal is described by a periodic
supercell with $10$ atoms, evenly spaced. We take the distance between
adjacent atoms to be $a=5.29\angstrom$. The one-particle Hamiltonian is given
by 
\begin{equation}
	\bvec{H}=-\dfrac{1}{2}\frac{\partial^2}{\partial x^2} + V	.
	\label{}
\end{equation}
The potential $V$ is given by a sum of Gaussians
centered at the atoms with width $\sigma = 1.32\angstrom$ and depth $V_0 =
13.6\eV$. The kinetic energy is discretized using a simple 3-point finite
difference formula, resulting in a Hamiltonian $\bvec{H}$ with a discrete
eigenvalue spectrum with lower and upper eigenvalues equal to
$\eps_{-}=6.76\eV$ and $\eps_{+}=1959\eV$, respectively. Various temperatures
from $1024\K$ to $32\K$ were tried.  \figref{fig:nmm1dfmm} reports the
linear-log graph of $n_{\mathrm{MM}}$, the number of matrix by matrix
multiplications needed to evaluate the density matrix using our FOE, versus
$\beta\Delta\eps$, with $\beta\Delta\eps$ plotted in a logarithmic scale.  The
logarithmic dependence can be clearly seen. The prefactor of the logarithmic
dependence is rather small: when $\beta\Delta\eps$ is doubled, a number of
additional matrix multiplications equal to $17$ is required to achieve
two-digit accuracy ($D=2$), a number equal to $19$ is needed for $D=4$, and a
number equal to $21$ is needed for $D=6$, respectively. The observed
$D$-dependence of the number of matrix multiplications agrees well with the
prediction in \eqref{eq:nmmmdetail}. 

In order to assess the validity of the criterion for the number of matrix
multiplications given in Eq. (23), we report in \tabref{tab:energydensity1d}
the calculated relative energy error and relative density error, respectively,
at different temperatures, when the number of matrix multiplications is
bounded as in formula (23) using different values for $D$. The relative energy
error, $\Delta\eps_{\mathrm{rel}}$, measures the accuracy in the calculation
of the total electronic energy corresponding to the supercell 
$E = \tr(\bvec{\rho}\bvec{H})$. It is defined as     
\begin{equation}
	\Delta\eps_{\mathrm{rel}}=\frac{\abs{\hat{E}-E}}{\abs{E}}.  \label{}
\end{equation}
Similarly the relative $L^1$ error in the density function in real space is defined as
\begin{equation}
	\Delta\rho_{\mathrm{rel}}=\frac{\tr{\abs{\bvec{\hat{\rho}}-
	\bvec{\rho}} } }{\tr{\bvec{\rho}}}.
	\label{}
\end{equation}
Because $\tr{\bvec{\rho}}=N_{\mathrm{e}}$, where $ N_{\mathrm{e}}$ is the total
number of electrons in the supercell, $\Delta\rho_{\mathrm{rel}}$ is the same
as the $L^1$ density error per electron.  Table~\ref{tab:energydensity1d}
shows that for all the values of $\beta\Delta\eps$, our algorithm gives a
numerical accuracy that is even better than the target accuracy $D$. This is
not surprising because our theoretical analysis was based on the most
conservative error estimates. 

\subsection{Periodic three dimensional tight-binding model}\label{sec:TestB}

In this example we consider a periodic three dimensional single-band
tight-binding Hamiltonian in a simple cubic lattice. The Hamiltonian, which
can be viewed as the discretized form of a free-particle Hamiltonian, is given
in second quantized notation by: 
\begin{equation}
	\bvec{H} = -t \sum_{<i,j>} c_i^+ c_j,
	\label{}
\end{equation}
where the sum includes the nearest neighbors only. Choosing a value of
$2.27\eV$ for the hopping parameter $t$ the band extrema occur at
$\eps_{+}=13.606\eV$, and at $\eps_{-}=-13.606\eV$, respectively. In the
numerical calculation we consider a periodically repeated supercell with
$1000$ sites and chose a value of $100K$ for the temperature.
\tabref{tab:free3d} shows the dependence of $n_{\mathrm{MM}},\
\Delta\eps_{\mathrm{rel}},$ and  $\Delta\rho_{\mathrm{rel}}$ on the chemical
potential $\mu$, for different $D$ choices. Compared to the previous one
dimensional example in which $\beta\Delta\eps$ was as large as $7.12\times
10^{5}$, here $\beta\Delta\eps=1600$ due to the much smaller spectral width of
the tight-binding Hamiltonian. When $\mu=0$ the chemical potential lies
exactly in the middle of the spectrum. This symmetry leads to a relative error
as low as $10^{-19}$ for the density function. 

\subsection{Three dimensional disordered Anderson model}\label{sec:TestC}

In this example we consider an Anderson model with on-site disorder on a
simple cubic lattice. The Hamiltonian is given by 
\begin{equation}
	\bvec{H} = -t \sum_{<i,j>} c_i^+ c_j + \sum_i \eps_i c_i^+ c_i.
	\label{}
\end{equation}
This Hamiltonian contains random
on-site energies $\eps_i$ uniformly distributed in the interval
$[-1.13\eV,1.13\eV]$, and we use the same hopping parameter $t$ as in the
previous (ordered) example. In the numerical calculation we consider, as
before, a supercell with $1000$ sites with periodic boundary conditions, and
choose again a temperature of $100\K$. In one realization of disorder
corresponding to a particular set of random on-site energies, the spectrum has
extrema at $\eps_{+}=13.619\eV$ and at $\eps_{-}=-13.676\eV$. The effect of
disorder on the density function is remarkable: while in the periodic
tight-binding case the density was uniform, having the same constant value at
all the lattice sites, now the density is a random function in the lattice
sites within the supercell. \tabref{tab:rand3d} reports for the disordered
model the same data that were reported in \tabref{tab:free3d} for the ordered
model. We see that the accuracy of our numerical FOE is the same in the two
cases, irrespective of disorder. The only difference is that the super
convergence due to symmetry for $\mu=0$ no longer exists in the disordered
case. 

\section{Conclusion}
We proposed a multipole representation for the Fermi operator.
Based on this expansion, a rather simple and efficient algorithm was developed for 
electronic structure analysis. We have shown that the number of
number of matrix by matrix multiplication that are needed scales as
$(\ln\beta\Delta\eps)\cdot(\ln\ln \beta\Delta\eps)$ with  very
small overhead. Numerical examples show that the algorithm is promising and
has the potential to be applied to metallic systems. 

We have only considered the number of matrix matrix multiplications
as a measure for the computational cost. The real operational count should 
of course take into account the cost of multiplying two matrices, and hence
depends on how the matrices are represented. 
This is work in progress.

\appendix

\section{Mittag-Leffler's theorem and pole expansion for hyperbolic tangent function}
To obtain the pole expansion for hyperbolic tangent function $\tanh(z)$, 
we need a special case of the general Mittag-Leffler's theorem 
on the expansions of meromorphic functions (see, for example, 
\cite{JeffreysJeffreys, Ahlfors}).
\begin{theorem}[Mittag-Leffler]\label{thm:mit}
If a function $f(z)$ analytic at the origin has no singularities other than
poles for finite $z$, and if we can choose a sequence of contours $C_m$ about
$z=0$ tending to infinity, such that $\abs{f(z)}\leq M$ on $C_m$ and 
$\int_{C_m} \abs{\ud z/z}$ is uniformly bounded, then we have
\begin{equation}
f(z) = f(0) + \lim_{m\to\infty}\{P_m(z)-P_m(0)\},
\end{equation}
where $P_m(z)$ is the sum of the principal parts of $f(z)$ at all poles within
$C_m$.
\end{theorem}

For $\tanh(z) = \dfrac{\exp(z)-\exp(-z)}{\exp(z)+\exp(-z)}$, 
it is analytic at the origin and $\tanh(0) = 0$. The function has simple 
poles at $z_l = (l-1/2)\pi i, \ l\in\ZZ$ with principle parts $(z-z_l)^{-1}$.
Let us take the contours as 
\begin{equation*}
C_m = \{x\pm im\pi\mid \abs{x} \leq m\pi\}\cup
\{\pm m\pi+iy\mid \abs{y} \leq m\pi\},
\quad m\in\ZZ_{+}, 
\end{equation*}
it is then easy to verify that $C_m$ satisfy the conditions in 
\thmref{thm:mit}. According to \thmref{thm:mit},
\begin{equation}
\tanh(z) = \tanh(0) + \lim_{m\to\infty}\sum_{l=-m+1}^m 
\Bigl(\frac{1}{z-z_l} + \frac{1}{z_l}\Bigr).
\end{equation}
By symmetry of $z_l$, the second term within the brackets cancels,
and we arrive at \eqref{eq:poleexpnc}.

\noindent{\bf Acknowledgement:}
We thank M. Ceriotti and M. Parrinello for useful discussions. This work was
partially supported by DOE under Contract No. DE-FG02-03ER25587 and by ONR
under Contract No. N00014-01-1-0674 (W. E, L. L., and J. L.), and by DOE under
Contract No. DE-FG02-05ER46201 and NSF-MRSEC Grant DMR-02B706 (R. C. and L. L.).

\bibliographystyle{apsrev}
\bibliography{metal}

\begin{thebibliography}{17}
\expandafter\ifx\csname natexlab\endcsname\relax\def\natexlab#1{#1}\fi
\expandafter\ifx\csname bibnamefont\endcsname\relax
  \def\bibnamefont#1{#1}\fi
\expandafter\ifx\csname bibfnamefont\endcsname\relax
  \def\bibfnamefont#1{#1}\fi
\expandafter\ifx\csname citenamefont\endcsname\relax
  \def\citenamefont#1{#1}\fi
\expandafter\ifx\csname url\endcsname\relax
  \def\url#1{\texttt{#1}}\fi
\expandafter\ifx\csname urlprefix\endcsname\relax\def\urlprefix{URL }\fi
\providecommand{\bibinfo}[2]{#2}
\providecommand{\eprint}[2][]{\url{#2}}

\bibitem[{\citenamefont{Baroni and Giannozzi}(1992)}]{BaroniGiannozzi1992}
\bibinfo{author}{\bibfnamefont{S.}~\bibnamefont{Baroni}} \bibnamefont{and}
  \bibinfo{author}{\bibfnamefont{P.}~\bibnamefont{Giannozzi}},
  \bibinfo{journal}{Europhys. Lett.} \textbf{\bibinfo{volume}{17}},
  \bibinfo{pages}{547} (\bibinfo{year}{1992}).

\bibitem[{\citenamefont{Goedecker and Colombo}(1994)}]{goedeckercolombo1994}
\bibinfo{author}{\bibfnamefont{S.}~\bibnamefont{Goedecker}} \bibnamefont{and}
  \bibinfo{author}{\bibfnamefont{L.}~\bibnamefont{Colombo}},
  \bibinfo{journal}{Phys. Rev. Lett.} \textbf{\bibinfo{volume}{73}},
  \bibinfo{pages}{122} (\bibinfo{year}{1994}).

\bibitem[{\citenamefont{Goedecker and Teter}(1995)}]{goedeckerteter1995}
\bibinfo{author}{\bibfnamefont{S.}~\bibnamefont{Goedecker}} \bibnamefont{and}
  \bibinfo{author}{\bibfnamefont{M.}~\bibnamefont{Teter}},
  \bibinfo{journal}{Phys. Rev. B} \textbf{\bibinfo{volume}{51}},
  \bibinfo{pages}{9455} (\bibinfo{year}{1995}).

\bibitem[{\citenamefont{Goedecker}(1999)}]{goedecker1999}
\bibinfo{author}{\bibfnamefont{S.}~\bibnamefont{Goedecker}},
  \bibinfo{journal}{Rev. Mod. Phys.} \textbf{\bibinfo{volume}{71}},
  \bibinfo{pages}{1085} (\bibinfo{year}{1999}).

\bibitem[{\citenamefont{Krajewski and Parrinello}(2005)}]{parrinello2005}
\bibinfo{author}{\bibfnamefont{F.}~\bibnamefont{Krajewski}} \bibnamefont{and}
  \bibinfo{author}{\bibfnamefont{M.}~\bibnamefont{Parrinello}},
  \bibinfo{journal}{Phys. Rev. B} \textbf{\bibinfo{volume}{71}},
  \bibinfo{eid}{233105} (\bibinfo{year}{2005}).

\bibitem[{\citenamefont{Krajewski and
  Parrinello}(2006{\natexlab{a}})}]{parrinello2006a}
\bibinfo{author}{\bibfnamefont{F.}~\bibnamefont{Krajewski}} \bibnamefont{and}
  \bibinfo{author}{\bibfnamefont{M.}~\bibnamefont{Parrinello}},
  \bibinfo{journal}{Phys. Rev. B} \textbf{\bibinfo{volume}{73}},
  \bibinfo{pages}{41105} (\bibinfo{year}{2006}{\natexlab{a}}).

\bibitem[{\citenamefont{Krajewski and
  Parrinello}(2006{\natexlab{b}})}]{parrinello2006b}
\bibinfo{author}{\bibfnamefont{F.}~\bibnamefont{Krajewski}} \bibnamefont{and}
  \bibinfo{author}{\bibfnamefont{M.}~\bibnamefont{Parrinello}},
  \bibinfo{journal}{Phys. Rev. B} \textbf{\bibinfo{volume}{74}},
  \bibinfo{pages}{125107} (\bibinfo{year}{2006}{\natexlab{b}}).

\bibitem[{\citenamefont{Krajewski and Parrinello}(2007)}]{parrinello2007}
\bibinfo{author}{\bibfnamefont{F.}~\bibnamefont{Krajewski}} \bibnamefont{and}
  \bibinfo{author}{\bibfnamefont{M.}~\bibnamefont{Parrinello}},
  \bibinfo{journal}{Phys. Rev. B} \textbf{\bibinfo{volume}{75}},
  \bibinfo{pages}{235108} (\bibinfo{year}{2007}).

\bibitem[{\citenamefont{Ceriotti
  et~al.}(2008{\natexlab{a}})\citenamefont{Ceriotti, K{\"u}hne, and
  Parrinello}}]{parrinello2008}
\bibinfo{author}{\bibfnamefont{M.}~\bibnamefont{Ceriotti}},
  \bibinfo{author}{\bibfnamefont{T.}~\bibnamefont{K{\"u}hne}},
  \bibnamefont{and}
  \bibinfo{author}{\bibfnamefont{M.}~\bibnamefont{Parrinello}},
  \bibinfo{journal}{J. Chem. Phys} \textbf{\bibinfo{volume}{129}},
  \bibinfo{pages}{024707} (\bibinfo{year}{2008}{\natexlab{a}}).

\bibitem[{\citenamefont{Ceriotti
  et~al.}(2008{\natexlab{b}})\citenamefont{Ceriotti, K{\"u}hne, and
  Parrinello}}]{parrinelloarxiv}
\bibinfo{author}{\bibfnamefont{M.}~\bibnamefont{Ceriotti}},
  \bibinfo{author}{\bibfnamefont{T.}~\bibnamefont{K{\"u}hne}},
  \bibnamefont{and}
  \bibinfo{author}{\bibfnamefont{M.}~\bibnamefont{Parrinello}},
  \bibinfo{journal}{arXiv:0809.2232v1}  (\bibinfo{year}{2008}{\natexlab{b}}).

\bibitem[{\citenamefont{Liang et~al.}(2003)\citenamefont{Liang, Saravanan,
  Shao, Baer, Bell, and Head-Gordon}}]{headgordon2003}
\bibinfo{author}{\bibfnamefont{W.}~\bibnamefont{Liang}},
  \bibinfo{author}{\bibfnamefont{C.}~\bibnamefont{Saravanan}},
  \bibinfo{author}{\bibfnamefont{Y.}~\bibnamefont{Shao}},
  \bibinfo{author}{\bibfnamefont{R.}~\bibnamefont{Baer}},
  \bibinfo{author}{\bibfnamefont{A.~T.} \bibnamefont{Bell}}, \bibnamefont{and}
  \bibinfo{author}{\bibfnamefont{M.}~\bibnamefont{Head-Gordon}},
  \bibinfo{journal}{J. Chem. Phys.} \textbf{\bibinfo{volume}{119}},
  \bibinfo{pages}{4117} (\bibinfo{year}{2003}).

\bibitem[{\citenamefont{Liang et~al.}(2004)\citenamefont{Liang, Baer,
  Saravanan, Shao, Bell, and Head-Gordon}}]{headgordon2004}
\bibinfo{author}{\bibfnamefont{W.}~\bibnamefont{Liang}},
  \bibinfo{author}{\bibfnamefont{R.}~\bibnamefont{Baer}},
  \bibinfo{author}{\bibfnamefont{C.}~\bibnamefont{Saravanan}},
  \bibinfo{author}{\bibfnamefont{Y.}~\bibnamefont{Shao}},
  \bibinfo{author}{\bibfnamefont{A.~T.} \bibnamefont{Bell}}, \bibnamefont{and}
  \bibinfo{author}{\bibfnamefont{M.}~\bibnamefont{Head-Gordon}},
  \bibinfo{journal}{J. Comput. Phys.} \textbf{\bibinfo{volume}{194}},
  \bibinfo{pages}{575 } (\bibinfo{year}{2004}).

\bibitem[{\citenamefont{Mahan}(2000)}]{mahan2000mpp}
\bibinfo{author}{\bibfnamefont{G.}~\bibnamefont{Mahan}},
  \emph{\bibinfo{title}{{Many-particle Physics}}} (\bibinfo{publisher}{Plenum
  Pub Corp}, \bibinfo{year}{2000}).

\bibitem[{\citenamefont{Greengard and Rokhlin}(1987)}]{GreengardRokhlin1987}
\bibinfo{author}{\bibfnamefont{L.}~\bibnamefont{Greengard}} \bibnamefont{and}
  \bibinfo{author}{\bibfnamefont{V.}~\bibnamefont{Rokhlin}},
  \bibinfo{journal}{J. Comput. Phys.} \textbf{\bibinfo{volume}{73}},
  \bibinfo{pages}{325} (\bibinfo{year}{1987}).

\bibitem[{\citenamefont{Jeffreys and Jeffreys}(1956)}]{JeffreysJeffreys}
\bibinfo{author}{\bibfnamefont{H.}~\bibnamefont{Jeffreys}} \bibnamefont{and}
  \bibinfo{author}{\bibfnamefont{B.~S.} \bibnamefont{Jeffreys}},
  \emph{\bibinfo{title}{Methods of mathematical physics}}
  (\bibinfo{publisher}{Cambridge, at the University Press},
  \bibinfo{year}{1956}), \bibinfo{edition}{3rd} ed.

\bibitem[{\citenamefont{Ahlfors}(1978)}]{Ahlfors}
\bibinfo{author}{\bibfnamefont{L.~V.} \bibnamefont{Ahlfors}},
  \emph{\bibinfo{title}{Complex analysis}} (\bibinfo{publisher}{McGraw-Hill
  Book Co.}, \bibinfo{address}{New York}, \bibinfo{year}{1978}),
  \bibinfo{edition}{3rd} ed.

\bibitem[{\citenamefont{Baer and Head-Gordon}(1997)}]{headgordon1997}
\bibinfo{author}{\bibfnamefont{R.}~\bibnamefont{Baer}} \bibnamefont{and}
  \bibinfo{author}{\bibfnamefont{M.}~\bibnamefont{Head-Gordon}},
  \bibinfo{journal}{J. Chem. Phys.} \textbf{\bibinfo{volume}{107}},
  \bibinfo{pages}{10003} (\bibinfo{year}{1997}).

\end{thebibliography}

\newpage

\begin{table}[ht]
	\centering
	\begin{tabular}{c|c||c|c|c|c|c|c}
		\toprule
		\multirow{2}*{T}&\multirow{2}*{$\beta\Delta\eps$} &\multicolumn{3}{c|}{$\Delta\eps_{\mathrm{rel}}$}
		&\multicolumn{3}{c}{$\Delta\rho_{\mathrm{rel}}$}\\
		\cmidrule{3-8}
		 & &$D=2$& $D=4$ & $D=6$ &$D=2$& $D=4$ & $D=6$\\
		\midrule
		$1024K$ & $2.22\times 10^{4}$ & $1.64\times 10^{-3}$ & $5.98\times 10^{-6}$ & $3.31\times 10^{-8}$ &
		$4.21\times 10^{-4}$ & $2.23\times 10^{-6}$ & $1.50\times 10^{-8}$\\
		$ 512K$ & $4.44\times 10^{4}$ & $1.73\times 10^{-3}$ & $6.49\times 10^{-6}$ & $3.70\times 10^{-8}$ &
		$4.63\times 10^{-4}$ & $2.52\times 10^{-6}$ & $1.74\times 10^{-8}$\\
		$ 256K$ & $8.89\times 10^{4}$ & $1.78\times 10^{-3}$ & $6.83\times 10^{-6}$ & $3.96\times 10^{-8}$ &
		$4.77\times 10^{-4}$ & $2.62\times 10^{-6}$ & $1.81\times 10^{-8}$\\
		$ 128K$ & $1.78\times 10^{5}$ & $1.74\times 10^{-3}$ & $6.55\times 10^{-6}$ & $3.75\times 10^{-8}$ &
		$5.04\times 10^{-4}$ & $2.80\times 10^{-6}$ & $1.95\times 10^{-8}$\\
		$  64K$ & $3.56\times 10^{5}$ & $1.75\times 10^{-3}$ & $6.62\times 10^{-6}$ & $3.80\times 10^{-8}$ &
		$4.92\times 10^{-4}$ & $2.70\times 10^{-6}$ & $1.86\times 10^{-8}$\\
		$  32K$ & $7.12\times 10^{5}$ & $1.76\times 10^{-3}$ & $6.66\times 10^{-6}$ & $3.82\times 10^{-8}$ &
		$4.84\times 10^{-4}$ & $2.64\times 10^{-6}$ & $1.80\times 10^{-8}$\\
		\bottomrule
	\end{tabular}
	\caption{One dimensional Hamiltonian model of \secref{sec:TestA}. Relative energy error $\Delta\eps_{\mathrm{rel}}$ and
	relative $L^1$ density error $\Delta\rho_{\mathrm{rel}}$ for
	a large range of values of $\beta\Delta\eps$ and several values of $D$.}
	\label{tab:energydensity1d}
\end{table}

\begin{table}[ht]
	\centering
	\begin{tabular}{c||c|c|c|c|c|c}
		\toprule
		\multirow{2}*{$\mu$}& \multicolumn{3}{c|}{$D=4$}
		&\multicolumn{3}{c}{$D=8$}\\
		\cmidrule{2-7}
		& $n_{\mathrm{MM}} $ & $\Delta\eps_{\mathrm{rel}}$ &
		$\Delta\rho_{\mathrm{rel}}$ & $n_{\mathrm{MM}} $ &
		$\Delta\eps_{\mathrm{rel}}$ & $\Delta\rho_{\mathrm{rel}}$\\
		\midrule
		$-10.88\eV$ & $320$ & $4.09\times 10^{-9}$ & $2.31\times 10^{-10}$ & $376$ & $2.27\times 10^{-13}$ &
		$2.37\times 10^{-14}$\\
		$-5.44\eV$ & $308$ & $1.48\times 10^{-9}$ & $3.15\times 10^{-11}$ & $356$ & $4.77\times 10^{-13}$ &
		$2.52\times 10^{-15}$\\
		$ 0.00\eV$ & $305$ & $1.55\times 10^{-9}$ & $6.26\times 10^{-19}$ & $357$ & $2.98\times 10^{-15}$ &
		$6.26\times 10^{-19}$\\
		$ 5.44\eV$ & $308$ & $1.45\times 10^{-8}$ & $1.34\times 10^{-12}$ & $356$ & $5.36\times 10^{-13}$ &
		$1.07\times 10^{-16}$\\
		$10.88\eV$ & $320$ & $1.69\times 10^{-8}$ & $1.78\times 10^{-13}$ & $376$ & $1.09\times 10^{-12}$ &
		$1.80\times 10^{-17}$\\
		\bottomrule
	\end{tabular}
	\caption{Three dimensional periodic tight binding model of
	\secref{sec:TestB}. Number of matrix matrix multiplications
	$n_{\mathrm{MM}}$, relative energy error $\Delta\eps_{\mathrm{rel}}$, and
	relative $L^1$ density error $\Delta\rho_{\mathrm{rel}}$. 
	 For $\mu=0$, the algorithm achieves machine accuracy for the absolute error
	 of the density function as a consequence of symmetry.} 
	 \label{tab:free3d}
\end{table}

\begin{table}[ht]
	\centering
	\begin{tabular}{c||c|c|c|c|c|c}
		\toprule
		\multirow{2}*{$\mu$}& \multicolumn{3}{c|}{$D=4$}
		&\multicolumn{3}{c}{$D=8$}\\
		\cmidrule{2-7}
		& $n_{\mathrm{MM}} $ & $\Delta\eps_{\mathrm{rel}}$ &
		$\Delta\rho_{\mathrm{rel}}$ & $n_{\mathrm{MM}} $ &
		$\Delta\eps_{\mathrm{rel}}$ & $\Delta\rho_{\mathrm{rel}}$\\
		\midrule
		$-10.88\eV$ & $320$ & $5.16\times 10^{-9}$ & $1.72\times 10^{-10}$ & $376$ & $3.16\times 10^{-13}$ &
		$2.59\times 10^{-14}$\\
		$-5.44\eV$ & $308$ & $4.75\times 10^{-9}$ & $2.43\times 10^{-11}$ & $356$ & $3.71\times 10^{-13}$ &
		$1.48\times 10^{-15}$\\
		$ 0.00\eV$ & $305$ & $8.08\times 10^{-10}$ & $9.50\times 10^{-13}$ & $357$ & $1.76\times 10^{-14}$ &
		$2.39\times 10^{-17}$\\
		$ 5.44\eV$ & $308$ & $1.01\times 10^{-8}$ & $1.22\times 10^{-12}$ & $356$ & $3.57\times 10^{-13}$ &
		$8.05\times 10^{-17}$\\
		$10.88\eV$ & $320$ & $1.30\times 10^{-8}$ & $1.56\times 10^{-13}$ & $376$ & $9.56\times 10^{-13}$ &
		$1.83\times 10^{-17}$\\
		\bottomrule
	\end{tabular}
	\caption{Three dimensional Anderson model with on-site disorder discussed in
	\secref{sec:TestC}. Number of matrix matrix multiplications
	$n_{\mathrm{MM}}$, relative energy error $\Delta\eps_{\mathrm{rel}}$, and
	relative $L^1$ density error $\Delta\rho_{\mathrm{rel}}$.}
	\label{tab:rand3d}
\end{table}

\end{document}